\documentstyle[preprint]{aastex}

\shortauthors{Burgasser et al.}
\shorttitle{T dwarf H{$\alpha$} Emission}

\begin{document}

\title{Detection of H{$\alpha$} Emission in a Methane (T-type) Brown Dwarf}

\author{Adam J. Burgasser\altaffilmark{1},
J. Davy Kirkpatrick\altaffilmark{2},
I. Neill Reid\altaffilmark{3},
James Liebert\altaffilmark{4},
John E. Gizis\altaffilmark{2},
and 
Michael E. Brown\altaffilmark{5,6}
}

\altaffiltext{1}{Division of Physics, M/S 103-33, 
California Institute of Technology, Pasadena, CA 91125; diver@its.caltech.edu}
\altaffiltext{2}{Infrared Processing and Analysis Center, M/S 100-22, 
California Institute of Technology, Pasadena, CA 91125; davy@ipac.caltech.edu, 
gizis@ipac.caltech.edu}
\altaffiltext{3}{Department of Physics and Astronomy, 
University of Pennsylvania, 209 South 33rd Street, Philadelphia, PA 19104-6396;
inr@herschel.physics.upenn.edu}
\altaffiltext{4}{Steward Observatory, University of Arizona,
Tucson, AZ 85721; liebert@as.arizona.edu}
\altaffiltext{5}{Division of Geological and Planetary Sciences, M/S 105-21, 
California Institute of Technology, Pasadena, California 91125; 
mbrown@gps.caltech.edu}
\altaffiltext{6}{Alfred P.\ Sloan Research Fellow}

\begin{abstract}

We report the detection of H{$\alpha$} emission in the T dwarf 
(methane brown dwarf) 2MASSW 
J1237392+652615 over three days using the Keck Low Resolution Imaging
Spectrograph.  The measured line flux, log(L$_{H{\alpha}}$/L$_{bol}$)
= $-$4.3, is roughly
consistent with early M dwarf activity levels and inconsistent with decreasing 
activity
trends in late M and L dwarfs.
Similar emission is not seen in
two other T dwarfs.  
We speculate on several mechanisms that may be responsible for
emission, 
including a strong magnetic field, continuous flaring, 
acoustic heat generation, and 
a close ($a$ $\sim$ 4 - 20 R$_{Jup}$)
interacting binary, with the cooler component overflowing its Roche lobe.
We suggest that the M9.5Ve PC 0025+0447 could be a warm analogue
to 2MASS J1237+65, and may be powered by the latter mechanism.
\end{abstract}


\section{Introduction}

Activity is an important parameter in the study of stellar populations.  Numerous
investigations of late-type (F-M) stars have shown correlations between 
emission 
(e.g., Ca II H \& K, Mg II h \& k, Balmer series, etc.) and fundamental 
 parameters such
as age, rotation, and metallicity \citep{Ha00}.  
It is generally believed that the majority of this
optical emission occurs in the chromosphere via collisional
heating by ions and electrons along magnetic field lines.  Indeed, 
this hypothesis is supported by the observed
correlation of activity and rotation in late-type stars \citep{Kr67,No84,Ba87},
which is expected if magnetic fields are generated by an internal dynamo
(e.g.\ $\alpha$-$\Omega$ dynamo; Parker 1955).  
Decrease in activity as stars age can be attributed to spin-down
due to angular momentum loss in stellar winds \citep{St86}.

As we examine cooler M and L dwarfs, however, these relations begin to break down.
As stars become fully convective ($\sim$ 0.3 M$_{\sun}$), the $\alpha$-$\Omega$ 
dynamo mechanism
becomes ineffective, as it requires a low-buoyancy, radiative/convective
boundary to anchor flux lines \citep{Sp80}. 
However, the observed activity level remains
roughly constant around this transition point \citep{Ha96}, 
suggesting a turbulent dynamo as an alternate magnetic field 
source \citep{Du93}.  Indeed, flaring activity, which is magnetically 
driven, is seen in objects as late as M9.5 \citep{Li99,Rd99}, 
supporting the existence
of substantial magnetic fields beyond the convective cut-off.
Alternately, acoustic heating could  
sufficiently heat the chromosphere \citep{Sc87,Md92} 
to produce a ``basal flux'' of H$\alpha$ emission.  In either case, 
\citet{Gi00} have shown that the fraction of objects with measurable
emission rises to 
100\% at spectral type M7, then rapidly declines, so that no emission is
seen in types L5 or later \citep{Ki00}.  
The decrease in (steady) activity even encompasses 
objects with rapid rotation \citep{Ba95,Ti98},
at odds with trends in hotter stars. 
Whether this drop in emission is due to ineffective chromospheric heating, decreased 
magnetic activity, or some other mechanism is unclear, but the end of the 
main sequence appears to mark a change in activity.

Based on the results of \citet{Gi00}, 
we would not expect significant activity in 
T dwarfs, brown dwarfs that show CH$_4$ absorption bands at 1.6 and
2.2 $\micron$ \citep{Ki99}.
Nonetheless, we have observed H$\alpha$ in emission in the T dwarf
2MASSW J1237392+652615 \citep[herefter 2MASS J1237+65]{Bg99}, 
identified from the Two Micron All Sky Survey \citep{Sk97}. 
We describe the optical observations of this and two other T dwarfs in 
$\S$2; in $\S$3 we discuss the H$\alpha$ detection in 2MASS J1237+65 
and possible emission mechanisms; 
in $\S$4 we compare 2MASS J1237+65 to the unusual M9.5Ve PC 0025+0447; 
we summarize our results in $\S$5.

\section{Optical Spectroscopy}

\subsection{Observations}
2MASS J1237+65 was observed on three consecutive nights, 1999 July 16, 17, and 18 
(UT) using the Low Resolution Imaging Spectrograph \citep[hereafter LRIS]{Ok95} 
on the Keck II 10-meter telescope. On each occasion, conditions were transparent
with seeing of 0{$\farcs$}8 to 1{$\farcs$}0, and we employed a 1$\arcsec$ wide
slit. The target was acquired via blind offset from field stars, as it was 
invisible in the guiding imager.  
The July 16 and 17 observations were made using the 400 lines mm$^{-1}$
grating blazed at 8500 {\AA}, covering the wavelength range 6300 to 10100 {\AA}
at 9 {\AA} resolution.
Total integrations
of 4800
(1800 + 1800 + 1200) and 1800 s were obtained, respectively.  
The July 17 observation was plagued by poor target centering, reducing the observed
flux by a factor of $\sim$ 2; the data for this night are thus
omitted, although H$\alpha$ was detected.  
The July 18 observation
was an 1800 s exposure using the 300 lines mm$^{-1}$ grating blazed at
5000 {\AA}, covering 3800 to 8600 {\AA} at 12 {\AA} resolution. 
Additional 3600 and 2700 s observations (total integration time) of 
T dwarfs SDSSp J162414.37+002915.6 \citep[hereafter SDSS 1624+00]{Ss99} 
and SDSSp J134646.45-003150.4 \citep[hereafter SDSS 1346-00]{Ts00} 
were made on July 16 and 17, using the 400 lines mm$^{-1}$ grating blazed at
8500 {\AA}.

The data for all three objects were reduced and calibrated using 
standard IRAF\footnote{IRAF is distributed 
by the National Optical Astronomy Observatories,
which are operated by the Association of Universities for Research
in Astronomy, Inc., under cooperative agreement with the National
Science Foundation.} routines. 
A one-second dark exposure was used to remove the bias, and quartz-lamp 
flat-field exposures were used to normalize the
response of the detector. The individual spectra were extracted 
using the APEXTRACT routine, allowing for slight curvature 
of the point-source dispersion line viewed through
the LRIS optics. 
Due to the low flux levels of the T dwarf targets,
we used an extraction template derived from observations of standard stars. 
Wavelength 
calibration was achieved using Hg-Ne-Ar arc lamp exposures taken 
after each object exposure. Finally, the spectra
were flux calibrated using observations of the DC 
white dwarf standard LTT 9491
\citep{Hm94}. Data have not 
been corrected for telluric absorption, so the
atmospheric H$_2$O bands at 
8161-8282, 8950-9300, and 9300-9650 {\AA} are still present in 
the spectra (telluric H$_2$O and O$_2$ bands are not visible shortward of
8000 {\AA} due to low flux levels).

\subsection{Optical Spectra}

Spectra from 6300-10100 {\AA} for all three objects are shown in Figure 1.
Data blueward
of 6300 {\AA} are not shown, as our July 18 observations failed to detect any
flux, continuum or emission, in this regime.
Features identified by
\citet{Op98} in Gl 229B are indicated.  
The optical spectrum of SDSS 1624+00 is discussed in detail in \citet{Li00}.
Cs I lines at 8521
and 8944 {\AA} are seen strongly in SDSS 1624+00, while 
both SDSS 1346-00 and 2MASS J1237+65 have 
progressively weaker lines.  
Broadened H$_2$O at 9250 {\AA} and a weak CH$_4$ feature
centered near 8950 {\AA} are noted, both of which are stronger in
2MASS J1237+65 and SDSS 1346-00 than in SDSS 1624+00.  
The pressure-broadened resonant 
K I doublet at 7665 and 7699 {\AA}, which is prominent in the spectra of L dwarfs
\citep{Ki99}, is masking most of the flux 
between 7300 and 8100 {\AA} in
all three objects \citep{Bu00, Li00}.  We note
a feature in SDSS 1624+00 centered around 
9900 {\AA} which we identify as the 9896 {\AA} 0-0 
A$^4$$\Delta$-X$^4$$\Delta$ band of FeH, a
feature that weakens from L6V and later \citep{Ki99}.  This detection supports 
the claims of \citet{Na00} and \citet{Li00} that SDSS 1624+00 is 
probably warmer than Gl 229B, which shows no FeH band.
Hydride bands in SDSS 1346-00 and 2MASS J1237+65 are either weak or absent,
suggesting that they in turn are cooler than SDSS 1624+00.  
The remaining feature of interest, H$\alpha$ emission at 
6563 {\AA} in 2MASS J1237+65, is the subject of the remainder of this article.
 
\placefigure{fig-1}

\section{H$\alpha$ in 2MASS J1237+65}

The H$\alpha$ emission seen in 2MASS J1237+65 (Figure 2) 
is quite unexpected given the 
cool nature of this object, and great care was taken to verify its
presence over three nights of observation.  The emission spike was clearly seen in
the raw data and is not spatially extended, thus
ruling out the possibility of a background HI region or telluric emission.  
H$\alpha$ emission was detected in five observations, spanning three nights
and two different instrument settings, with the emission full-width at 
half-maximum equivalent to the instrumental resolution ($\sim$ 5 pixels).
Hence, we can rule out the possibility of
a chance cosmic ray.  Finally, since no emission at 6563 {\AA} is seen for
the other T dwarfs, which were reduced using the same bias and flat-field
exposures, we can rule out detector features.  The H$\alpha$ emission line is thus
quite real.

Measurements of the H$\alpha$ line for all three T dwarfs are given in Table 1.
The integrated line luminosity 
at 6563 {\AA} averaged over the three July 16
observations is $f_{H{\alpha}}$ = 
(6.6$\pm$0.6) $\times$ 10$^{-17}$ erg cm$^{-2}$ s$^{-1}$.  
An equivalent width measurement is not possible as no continuum flux
was detected in this spectral region.
Note that there appears to be an increase in H$\alpha$
flux on July 18 to $f_{H{\alpha}}$ =
(8.5$\pm$0.3) $\times$ 10$^{-17}$ erg cm$^{-2}$ s$^{-1}$.
We estimate the bolometric flux of 2MASS J1237+65 using the 2MASS 
J-band magnitude of 15.90$\pm$0.06 and assuming a Gl 229B  
bolometric correction of BC$_J$ = 2.3 \citep{Mw96}.  This 
yields m$_{bol}$ = 18.2 and f$_{bol}$ $\approx$
1.3 $\times$ 10$^{-12}$ erg cm$^{-2}$ s$^{-1}$; 
thus, log($f_{H{\alpha}}$/$f_{bol}$) = 
log(L$_{H{\alpha}}$/L$_{bol}$) $\approx$ $-$4.3.  A limit 
on H$\beta$ emission at
4861 {\AA} is estimated at $f_{H{\beta}}$ $\lesssim$ 
2.0 $\times$ 10$^{-17}$ erg cm$^{-2}$ s$^{-1}$, resulting in a 
Balmer emission ratio on 1999 July 18 (UT) of
$f_{H{\beta}}$/$f_{H{\alpha}}$ $\lesssim$ 0.24.

\placefigure{fig-2}

\subsection{Unusually Strong Magnetic Activity}

The detection of H$\alpha$ is clearly inconsistent with the emission trend 
seen in objects later than L5V and indeed in brown dwarfs in general \citep{Gi00}.
The level of activity for 2MASS J1237+65 can be placed in context with other 
late-type emission-line objects by comparing relative flux 
emitted through the H$\alpha$ line.
\citet{Ha96} find a mean log(L$_{H{\alpha}}$/L$_{bol}$) $\approx$ $-$3.8 for 
field dMe stars with M$_{bol}$ $<$ 12 (spectral type M5V or earlier), while 
\citet{Gi00}
show that this falls to below $-$5 for L dwarfs.  Hence, even the activity
level of 2MASS J1237+65 is inconsistent with these trends, while  
SDSS 1346-00 and SDSS 1624+00 have upper limits of
log(L$_{H{\alpha}}$/L$_{bol}$) $\approx$
$-$5.3 and $-$5.7, respectively.  Does 2MASS J1237+65 have an unusually 
strong magnetic field, perhaps
powered by a dynamo mechanism different than that of 
M and L dwarfs?  Further 
examination of activity statistics in this cool regime are clearly needed.

\subsection{Flaring}

While sustained activity may appear to diminish toward later spectral types,
flaring does not.    
A good example of this is BRI 0021-0214, a rapid M9.5V
rotator \citep{Ba95} which in quiescence shows little or no H$\alpha$ emission 
\citep{Ba95,Ti97,Ti98}, yet was seen to flare by \citet{Rd99} with 
log(L$_{H{\alpha}}$/L$_{bol}$) = $-$4.2, similar to 2MASS J1237+65.  
The BRI 0021-0214 flare also showed no continuum emission, again
similar to the activity in 2MASS J1237+65. 
While flaring has not yet been observed in L dwarfs, the
occurrence of flares in the latest M dwarfs \citep{Rd99,Li99,Fl00} 
suggests that they may simply have not yet been observed.
However, most strong M-dwarf flares persist only a few
hours \citep{Ha91}, while smaller flares (which would show weaker continuum
emission) have timescales on the order of minutes \citep{Ne86}.
Thus, prolonged emission of 2MASS J1237+65
requires either a continuous flaring mechanism or 
fortuitous timing on our part.  Further observations of the emission line
are warranted to investigate more fully
the temporal behavior of this emission.

\subsection{Acoustic Flux}

\citet{Sc87} first pointed out that
a minimum ``basal flux'' seen in late-type stars could be attributed to 
chromospheric heating from acoustic waves.  We can estimate the amount of
acoustic flux from 2MASS J1237+65 by extrapolating the models of 
\citet{Ul96},
which improved on pivotal work done by \citet{Bo84}.  
Estimating
T$_{eff}$ $\approx$ 1000 K, log g(cm s$^{-2}$) $\approx$ 5, and using the 
scaling F$_{acoustic}$ $\sim$ T$_{eff}$$^{20}$ from the coolest points in 
their Table 1, we 
estimate F$_{acoustic}$ $\approx$ 4 $\times$ 10$^{-8}$ erg cm$^{-2}$ s$^{-1}$, or 
log(L$_{acoustic}$/L$_{bol}$) 
$\approx$ $-$15, well below the observed activity level.  
As such, acoustic energy
is not likely the source of activity in 2MASS J1237+65.

\subsection{An Interacting Binary}

\placefigure{fig-3}

Finally, it is possible that 2MASS J1237+65 is active as a result of a 
binary interaction with an equal magnitude or fainter companion.
The binarity of brown dwarfs is well established,
\citep{Ma98,Ba99,Ko99}, and the identification of three equal magnitude binaries
in an L dwarf sample of ten by \citet{Ko99} is
consistent with the binary fraction seen in late main sequence stars 
($\sim$ 35\%; Fischer \& Marcy 1992; Henry \& McCarthy 1993).  
Hence, it is reasonable to consider
that 2MASS J1237+65 could itself be double.  In this case, the companion 
would have to be equal-mass or smaller, as the optical continuum shows no
evidence of a warmer component nor any significant variation over three days.

An unusual feature of 
brown dwarf binaries is the possibility
of sustained Roche lobe overflow.  Because of its degenerate interior,
if a brown dwarf loses mass on a 
dynamical timescale ($\tau$ $\sim$ ($\frac{R^3}{GM})^{\frac{1}{2}}$ $\sim$ 1 hr), 
its radius increases approximately as
dlnR/dlnM $\approx$ -$\frac{1}{3}$ when its mass lies in the range 
5 - 70 M$_{Jup}$\footnote{M$_{Jup}$ = 
1.9 $\times$ 10$^{30}$ g = 0.00095 M$_{\sun}$,
R$_{Jup}$ = 7.1 $\times$ 10$^9$ cm = 0.10 R$_{\sun}$ \citep{Al73}.} \citep{Bu93}.  
Sustained conservative mass transfer then
occurs if the change in the Roche lobe of the secondary satisfies

\begin{equation}
\frac{dlnR_L}{dlnM_2} = \frac{8}{3}q - \frac{4}{3} - 
(q+1)\frac{0.4q^{\frac{2}{3}} + \frac{1}{3}q^{\frac{1}{3}}(1+q^{\frac{1}{3}})^{-1}}
{0.6q^{\frac{2}{3}} + ln(1+q^{\frac{1}{3}})} < -\frac{1}{3}.
\end{equation}

\noindent We have taken $q$ = $\frac{M_2}{M_1}$ as the mass fraction, $M_1$ and $M_2$ the primary
and secondary masses, and $R_L$ the secondary Roche lobe radius, as approximated
by \citet{Eg83}

\begin{equation}
R_L = a\frac{0.49q^{\frac{2}{3}}}{0.6q^{\frac{2}{3}} + ln(1+q^{\frac{1}{3}})}.
\end{equation}

\noindent Here $a$ is the binary separation.  
Equation (1) is satisfied when $q$ $<$ 0.63.  We have
calculated possible brown dwarf binary separations, periods, and edge-on
primary radial velocities as a function of $q$ (Figure 3) for primary masses 
$M_1$ = 30 and 70 M$_{Jup}$ and \citep{Zp69}

\begin{equation}
R_L = R_2 = \frac{{\pi}M_{(Jup)}^{-\frac{1}{3}}}{(1+1.8M_{(Jup)}^{-\frac{1}{2}})^{\frac{4}{3}}} R_{(Jup)}.
\end{equation}

\noindent M$_{(Jup)}$ and R$_{(Jup)}$ are the mass and radius of the secondary
in units normalized to Jupiter.  
This rough model shows that sustained outflow requires a separation
$a$ $\lesssim$ 4 - 20 R$_{Jup}$ (Figure 3a).  This close separation
may seem problematic; however, the brown dwarf spectroscopic binary PPl 15 
has been shown to have an orbital separation of about 60 R$_{Jup}$ 
\citep{Ba99}.  
Note that this scenario results in at least partial 
eclipsing for
orbital inclination $i$ $\lesssim$ 6 - 30$\degr$, 
suggesting that photometric monitoring
could detect a transit event in a period of $\sim$ 1 - 10 hours. 
Alternatively, radial
velocity monitoring of the H$\alpha$ line can produce a measurable variation
(v$_1$ $\approx$ 10 - 90 cos($i$) km s$^{-1}$) over this period.  

The emission mechanism 
from overflow accretion is 
uncertain, as a hot accretion disk is ruled out by the lack of thermal 
continuum.  Ionization along a shock front or magnetic streaming onto
the primary's pole are possibilities.  
Nonetheless, we find this an intriguing and, 
more importantly, observationally
constrainable hypothesis.

\section{PC 0025+0447: An Analogue?}

A possible analogue to 2MASS J1237+65 is the much warmer M9.5Ve PC 0025+0447,
which was identified by \citet{Sh91} in a search for high redshift
quasars due to its unusually strong Balmer line emission.  The contribution of 
H$\alpha$
luminosity to total luminosity (log(L$_{H{\alpha}}$/L$_{bol}$) = $-$3.4;
Schneider et al.\ 1991)
is a full order of magnitude stronger in this 
object than 2MASS J1237+65, and appears to be
steady over a timescale of at least 7 - 8 years with no
indication of flaring activity \citep{Sh91,Mo94,Ma99}.  
Both \citet{Sh91} and \citet{Ma99}
argue that PC 0025+0447 
is a young (600 Myr) brown dwarf; if this were the case, then 
it is possible that 2MASS J1237+65
is an evolved version of this object that has slowly 
declined in activity with age.  \citet{Ma99} propose
various mechanisms for the activity in PC 0025+0447, including sustained
flaring and emission from a highly active, low-density corona.  
\citet{Mo94}, however, argue against PC 0025+0447 being a brown dwarf based on
significant Li depletion, and propose that it is a Hyades-age ($\sim$ 
700 Myr) main sequence star with normal 
photospheric thermal emission.  

We propose a scenario in which PC 0025+0447 is itself an interacting binary,
consisting of a $\sim$ 0.1 M$_{\sun}$ M dwarf and a brown dwarf
companion losing mass to the primary.  This scenario
provides a natural explanation for the observed phenomena:
Roche lobe overflow from the companion provides a steady 
H$\alpha$ emission source, accretion may lead to the observed
variable veiling of the M dwarf
spectrum \citep{Ma99}, 
and the accreted material from the brown dwarf companion
may retain lithium at primordial abundance, leading to the intermittent
appearance of Li 6707 {\AA} absorption seen by \citet{Ma99},
without requiring
modification of the measured trigonometric parallax. 
Extrapolation from Figure 3 leads to similar maximal separations (3 - 25
R$_{Jup}$) and periods (1 - 14 hours) as for 2MASS J1237+65.
Unfortunately,
despite many photometric and spectroscopic observations of PC 0025+0447,
no time-resolved data is available to constrain this hypothesis.   
More detailed
modeling is required to determine the feasibility of these mechanisms.

\section{Summary}

We have reported the detection of H$\alpha$ emission in the T dwarf 
2MASS J1237+65, at a level of log(L$_{H{\alpha}}$/L$_{bol}$) = $-$4.3.
This emission is intriguing, as it is a salient exception to the
cool dwarf temperature-activity relations identified
thus far.  We have proposed 
various activity mechanisms, including a strong magnetic field, 
continuous flaring, and an 
interacting brown dwarf binary, but these are speculative guesses at best.
Comparison can be drawn with the M9.5Ve PC 0025+0447, which, if it is
a brown dwarf, could be a warm analogue to 2MASS J1237+65.  
Both objects could also be close binary systems 
with lower-mass brown dwarf companions, which are steadily transferring mass 
to their primaries by Roche lobe overflow.
Nonetheless, the mechanism for both objects remains unclear.
Further investigation of the temporal stability of the H$\alpha$ line 
in 2MASS J1237+65 and
searches for emission in other T dwarfs is clearly warranted.

\acknowledgements
A.\ J.\ B.\ acknowledges the assistance of Terry Stickel and Wayne Wack 
at the telescope, useful discussions with 
P.\ Goldreich and M.\ Marley, helpful comments from our referee S.\ Hawley,
and the foresight of S.\ Phinney for assigning the brown dwarf
binary problem in class (equation 1).
A.\ J.\ B., J.\ D.\ K., I.\ N.\ R., J.\ E.\ G., 
and J.\ L.\ acknowledge funding through a NASA/JPL grant to 2MASS
Core Project science. 
A.\ J.\ B., J.\ D.\ K., and J.\ E.\ G.\ acknowledge the support of the Jet Propulsion
Laboratory, California Institute of Technology, which is operated under
contract with the National Aeronautics and Space Administration.
Data presented herein were obtained at the W.\ M.\ Keck Observatory
which is operated as a scientific partnership among the California Institute of
Technology, the University of California, and the National Aeronautics and Space
Administration.  The Observatory was made possible by generous financial 
support of the W.\ M.\ Keck Foundation.

\figcaption[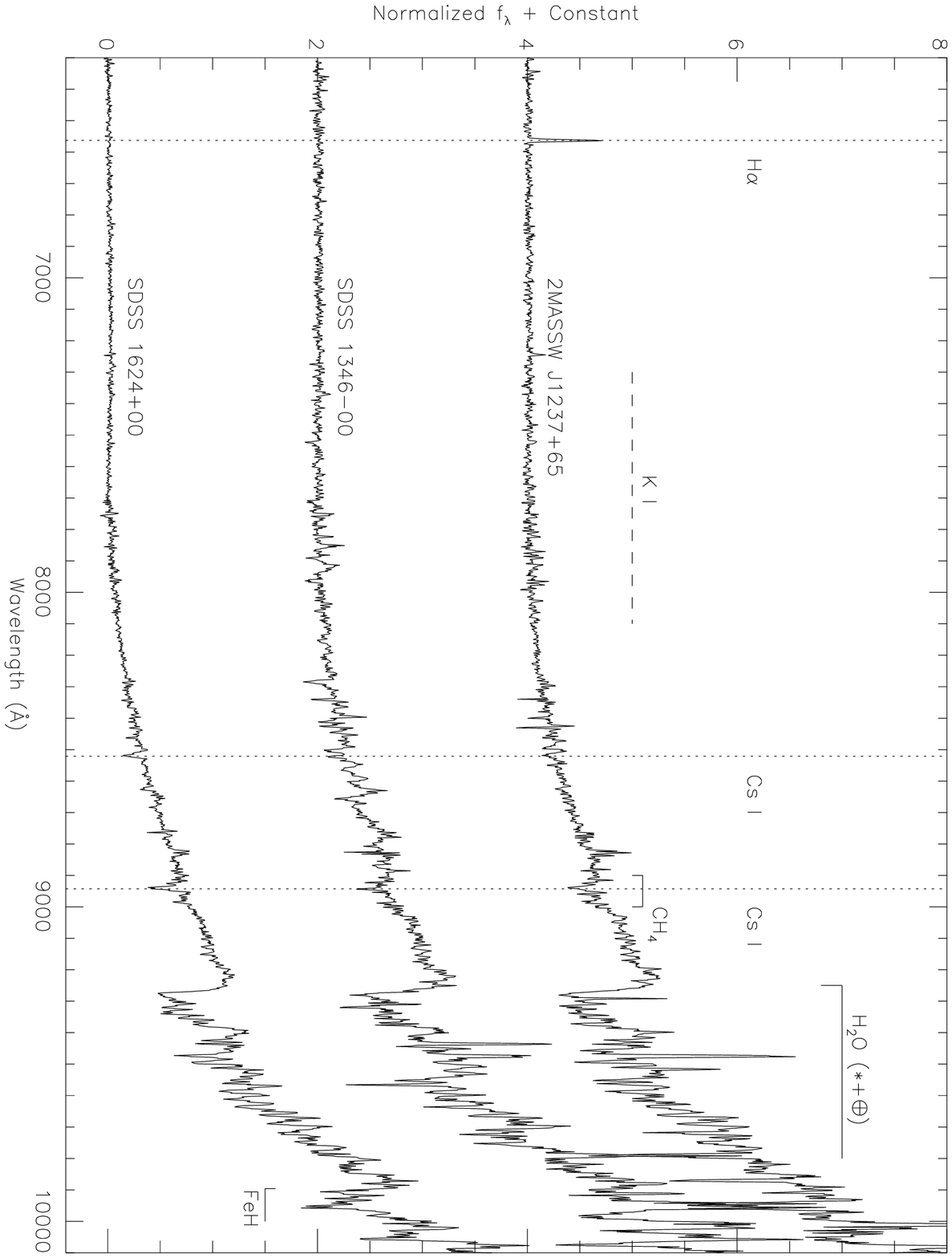]{LRIS optical spectra of three T dwarfs, 
2MASS J1237+65, SDSS 1346-00, and SDSS 1624+00, normalized at 9200 {\AA}.  
Data for SDSS 1346-00 and 2MASS J1237+65 have been offset vertically by 
constants of 2 and 4, respectively.  Common features ---
the K I resonant doublet (broadened to $\sim$ 7300-8100 {\AA}), 
Cs I (8521 and 8943 {\AA}), CH$_4$ (8950 {\AA}), and H$_2$O (9250 {\AA}, stellar
and telluric) --- are indicated, as is the 0-0 band of FeH (9896 {\AA})
in SDSS 1624+00.  
H$\alpha$ at 6563 {\AA} is 
unambiguously seen in 2MASS J1237+65. \label{fig-1}}

\figcaption[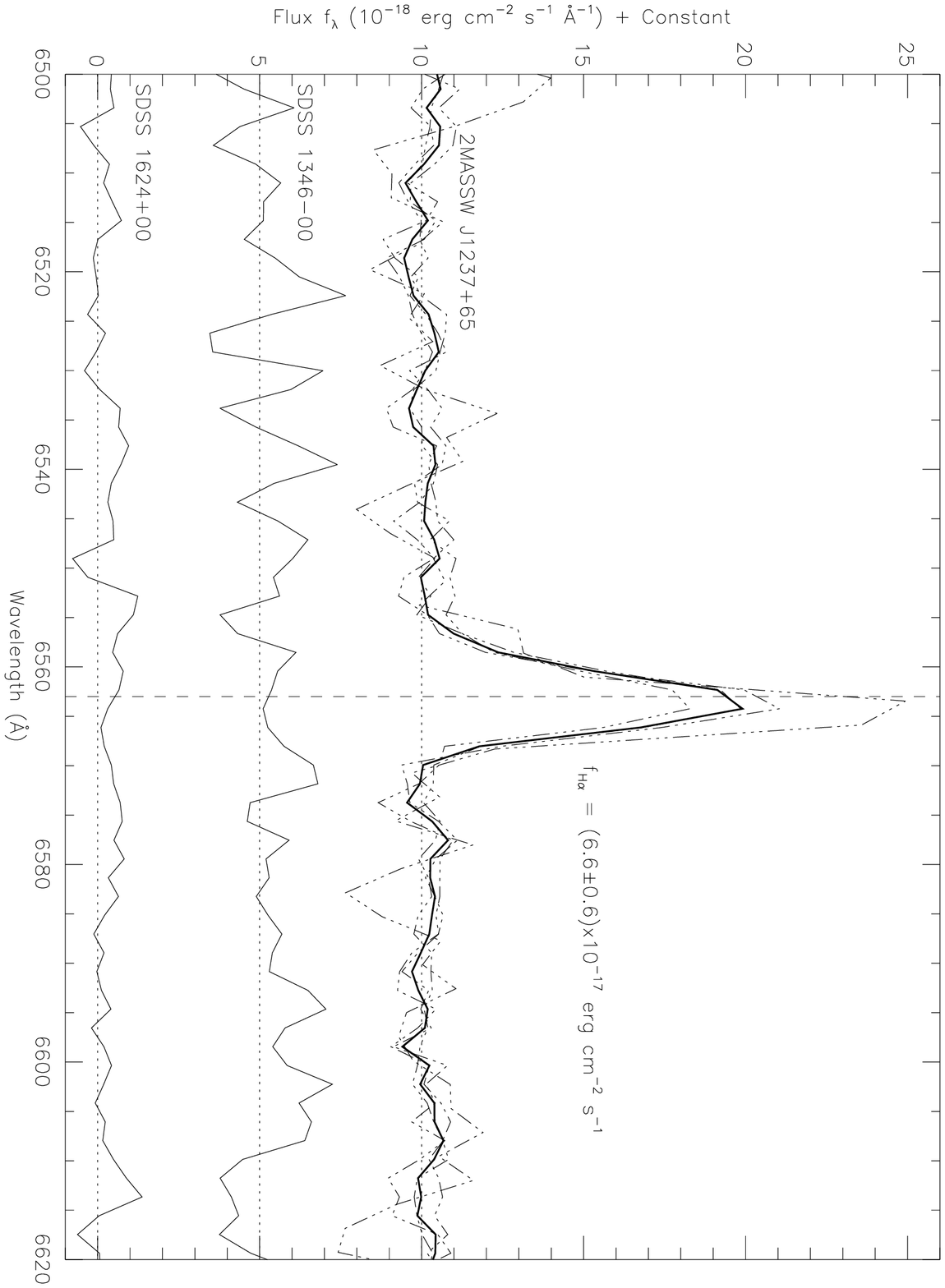]{A close-up of the H$\alpha$ feature in 2MASS J1237+65.
Four sets of observations for 2MASS J1237+65 are shown in dash-dotted lines,
while the mean from 1999 July 16 (UT) is shown as a solid thick line.  Data for
SDSS 1346-00 and SDSS 1624+00 are also shown.  Zero levels are indicated by
dotted lines, with SDSS 1346-00 and 2MASS J1237+65 offset vertically by 
constants of 5 $\times$ 10$^{-18}$ and 
10 $\times$ 10$^{-18}$ ergs cm$^{-2}$ s$^{-1}$ {\AA}$^{-1}$, respectively.
\label{fig-2}}

\figcaption[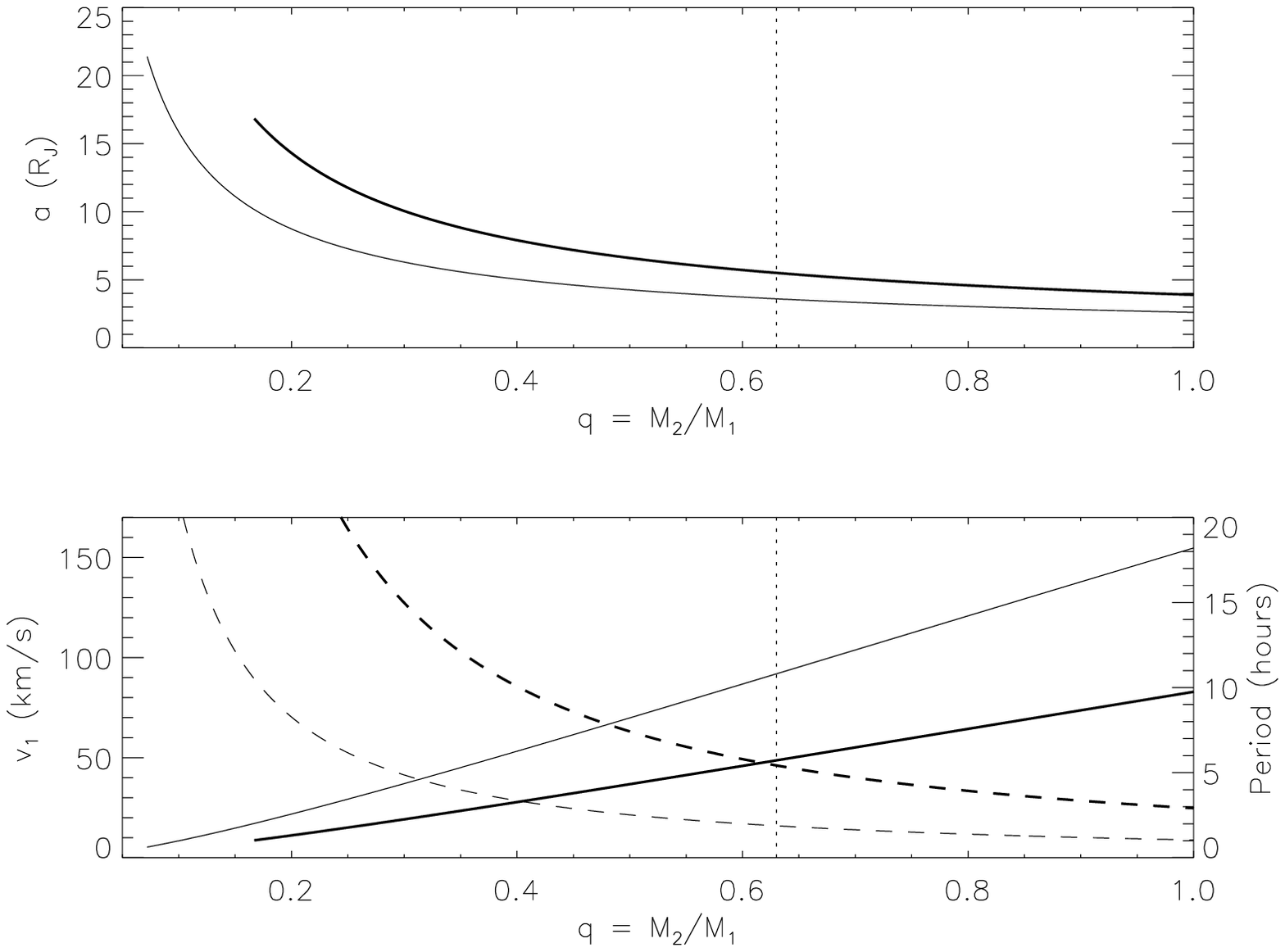]{(a) Orbital separation ($a$) versus mass ratio
($q$ = $\frac{M_2}{M_1}$) for a brown dwarf binary.  Lines indicate the
maximum separation in which the secondary
mass fills its Roche lobe for primary masses of 30 M$_{Jup}$ (thick line)
and 70 M$_{Jup}$ (thin line).  The dotted boundary indicates the maximum
$q$ = 0.63 required for sustained mass loss. (b) Maximum primary
radial velocities (solid lines) and orbital periods (dashed lines) for a Roche
lobe-filling secondary, for primary masses of 30 M$_{Jup}$ (thick line)
and 70 M$_{Jup}$ (thin line).  The limit $q$ = 0.63 is indicated as above.
\label{fig-3}}

\begin{deluxetable}{cccccccc}
\tabletypesize{\scriptsize}
\tablenum{1}
\tablewidth{0pt}
\tablecaption{Emission Measurements for Three T dwarfs. \label{tbl-1}}

\tablehead{
\colhead{} &
\multicolumn{3}{c}{Observation (UT)} &
\colhead{} &
\colhead{} &
\colhead{} &
 \\
\colhead{Object} &
\colhead{Date} &
\colhead{Time} &
\colhead{t$_{int}$ (s)} &
\colhead{$f_{H{\alpha}}$\tablenotemark{a}} &
\colhead{log(L$_{H{\alpha}}$/L$_{bol}$)\tablenotemark{b}} &
\colhead{$f_{H{\beta}}$\tablenotemark{a}} &
\colhead{log(L$_{H{\beta}}$/L$_{bol}$)\tablenotemark{b}} \\
}
\startdata
2MASS J1237+65 & 16 July 1999 & 06:21 & 1200 & 5.95$\pm$0.13 & $-$4.3 & -- & -- \\
2MASS J1237+65 & 16 July 1999 & 06:40 & 1800 & 7.06$\pm$0.09 & $-$4.3 & -- & -- \\
2MASS J1237+65 & 16 July 1999 & 07:15 & 1800 & 6.76$\pm$0.04 & $-$4.3 & -- & -- \\
SDSS 1624+00 & 16 July 1999 & 08:27 & 3600 & $<$ 0.4\tablenotemark{d} & $<$ $-$5.7 & -- & -- \\
SDSS 1346$-$00 & 17 July 1999 & 07:00 & 2700 & $<$ 0.7\tablenotemark{d} & $<$ $-$5.3 & -- & -- \\
2MASS J1237+65\tablenotemark{c} & 17 July 1999 & 07:56 & 1800 & 1.68$\pm$0.08 & $-$4.9 & -- & -- \\
2MASS J1237+65 & 18 July 1999 & 06:39 & 1800 & 8.50$\pm$0.30 & $-$4.2 & $<$ 2.0\tablenotemark{d} & $<$ $-$4.8\\
\tablenotetext{a}{In units of 10$^{-17}$ erg cm$^{-2}$ s$^{-1}$}
\tablenotetext{b}{Assuming BC$_J$ = 2.3 \citep{Mw96}.  J-band magnitudes
for the Sloan objects taken from \citet{Ss99} and \citet{Ts00}.}
\tablenotetext{c}{The object appears to have been improperly placed in the
slit for this observation, so that the entire spectrum is depressed by a 
factor of $\approx$ 2.  Data are included only to show detection of 
H$\alpha$.}
\tablenotetext{d}{Upper limits based on the continuum noise and spectral 
resolution of 9 {\AA} (H$\alpha$) and 12 {\AA} (H$\beta$).}
\enddata
\end{deluxetable}

\plotone{Burgasser.fig1.ps}

\plotone{Burgasser.fig2.ps}

\plotone{Burgasser.fig3.ps}

\end{document}